# What Makes Agile Software Development Agile?

Marco Kuhrmann, Paolo Tell, Regina Hebig, Jil Klünder, Jürgen Münch, Oliver Linssen, Dietmar Pfahl, Michael Felderer, Christian R. Prause, Stephen G. MacDonell, Joyce Nakatumba-Nabende, David Raffo, Sarah Beecham, Eray Tüzün, Gustavo López, Nicolas Paez, Diego Fontdevila, Sherlock A. Licorish, Steffen Küpper, Günther Ruhe, Eric Knauss, Özden Özcan-Top, Paul Clarke, Fergal McCaffery, Marcela Genero, Aurora Vizcaino, Mario Piattini, Marcos Kalinowski, Tayana Conte, Rafael Prikladnicki, Stephan Krusche, Ahmet Coşkunçay, Ezequiel Scott, Fabio Calefato, Svetlana Pimonova, Rolf-Helge Pfeiffer, Ulrik Pagh Schultz, Rogardt Heldal, Masud Fazal-Baqaie, Craig Anslow, Maleknaz Nayebi, Kurt Schneider, Stefan Sauer, Dietmar Winkler, Stefan Biffl, Maria Cecilia Bastarrica, and Ita Richardson

**Abstract**

*Together with many success stories, promises such as the increase in production speed and the improvement in stakeholders' collaboration have contributed to making agile a transformation in the software industry in which many companies want to take part. However, driven either by a natural and expected evolution or by contextual factors that challenge the adoption of agile methods as prescribed by their creator(s), software processes in practice mutate into hybrids over time. Are these still agile? In this article, we investigate the question: what makes a software development method agile? We present an empirical study grounded in a large-scale international survey that aims to identify software development methods and practices that improve or tame agility. Based on 556 data points, we analyze the perceived degree of agility in the implementation of standard project disciplines and its relation to used development methods and practices. Our findings suggest that only a small number of participants operate their projects in a purely traditional or agile manner (under 15%). That said, most project disciplines and most practices show a clear trend towards increasing degrees of agility. Compared to the methods used to develop software, the selection of practices has a stronger effect on the degree of agility of a given discipline. Finally, there are no methods or practices that explicitly guarantee or prevent agility. We conclude that agility cannot be defined solely at the process level. Additional factors need to be taken into account when trying to implement or improve agility in a software company. Finally, we discuss the field of software process-related research in the light of our findings and present a roadmap for future research.*

## 1. INTRODUCTION

Faster, higher, stronger—the Olympic motto[1] could also be the motto for today's software development practice. Software development needs to be creative, conducted by self-organizing cross-functional teams. Collaboration is key and working software is in the spotlight. This aspiration, which is also reflected in the "Agile Manifesto" [1], has become an ideal pursued by many companies. *Agile software development* has now been around for 20 years, and there is no denying that it has led to several improvements [2] such as increased speed of software development and intensified collaboration between different stakeholders.

However, when asking project managers and developers what agile software development means, the answer is likely: *Scrum* or *XP* [3]. Refining the question and adding further contextual factors, e.g., globally distributed software development or software development in regulated do- mains, the answer is not that simple anymore [4]–[6]. Due to the hype and numerous (partially) contradicting definitions, there is much confusion regarding the terminology and the concepts. However, quite often, people only *think* that they work agile or even pretend to work agile [3]. That is, there are software development processes that are *associated* with the term "agile software development", and there are methods and practices that are *perceived* as "agile". However, the use of many of these methods in practice is limited by contextual factors and in certain settings, practices often associated with agile may have been in use prior to the agile manifesto [7]. In response, projects compose their individual processes to address the respective needs (we refer to these as *hybrid methods* [8]). Still, companies want to participate in the "Agile Transformation" for various reasons and, therefore, there is also interest in creating "agile" methods.

---
[1] from Latin: "Citius, Altius, Fortius"

This motivates the primary question of this article: *What makes a software development method agile?*

In this article, we present an empirical study grounded in a large-scale international survey[2], which identified methods and practices that either enable or impede agility of software development methods. Based on 556 data points, we analyze the degree of agility in the implementation of eleven standard project disciplines, which are based on the SWEBOK categories [9], such as requirements engineering and testing as *perceived* by the survey participants. We link these reports to the 24 methods and 36 practices used by the participants to run their projects and we study if methods and practices affect the degree of agility.

Our findings indicate that few participants run their projects in a purely agile or a purely traditional manner, and most of them use home-grown hybrid development methods. Out of 660 cases (pairwise combinations of 60 methods and practices with 11 project disciplines), 146 show a significant shift in the perceived degree of agility when using specific methods and practices. Of these 146 changed perceptions, 88 cases show a trend of moving from traditional to agile. Another 29 cases show a significant shift with no explicit tendency towards agile or traditional. Hence, we conclude that about half of the shifts found are towards agile, even though we acknowledge that some shifts exist that are neutral regarding the perceived degree of agility.

The remainder of this article is organized as follows: Section 2 presents related work. Section 3 describes the research design before we present the results in Section 4. Section 5 provides a summary and a discussion of our findings. Furthermore, we derive a roadmap to steer future research before we conclude the paper in Section 6.

## 2. RELATED WORK

The development of software process models dates back to the 1960s and, over time, numerous different approaches have been published. From the very beginning, the different approaches used to organize software development have been critically discussed. These discussions have begun with Royce [10] stating that strict sequential models *do not*[3] properly reflect the needs of software development. In response, shorter iterations and incremental approaches became popular, such as Boehm's Spiral model [11] representing an iterative, risk-driven process, and Mills [12] and Dyer [13] suggesting incremental development processes in the 1970s and 1980s. In the following years, more software processes emerged. They included more methods, practices, and tools, becoming increasingly "heavy-weighted", for example the Rational Unified Process (RUP; [14]). Over time, developers began to reject these approaches as being too large, with too few degrees of freedom, or too much focus on documentation rather than on producing working software. A counter-movement started to move away from documentation- and specification-based software development towards *making software*, which culminated in Beck's Extreme Programming [15] and, eventually, in the Agile Manifesto in 2001 [1].

Complementing all these developments in practice, a considerable body of research has been accumulated on the software process. However, since researchers are naturally interested in "modern" software development [2], especially in the past two decades, research has been overloaded with results regarding drivers, challenges, benefits, practices, case studies, popularity and so on of agile processes (e.g., [2], [16]–[20]). This generates the impression that the "traditional" processes have been entirely replaced, only playing a role in process modeling in domains with special requirements (e.g., regulations and norms), or in discussions why certain companies do not use agile methods [19], [21], [22]. This trend can also be observed in many surveys that aim to collect evidence on agile methods, but barely put it into context. For instance, the *State of Agile* survey [23] on a global scale, and several regional surveys such as the *Swiss Agile Study* [24], the *Status Quo Agile* study [25], and [26] provides a snapshot of the status of agile methods being used in industry. However, these studies cover traditional processes only marginally—if at all. Furthermore, the perception of what "agile" actually means is fairly diffuse. For instance, Jalali et al. [27] conclude that agility is judged differently and is not a well-defined concept. This is also reflected in the considerable effort that is spent on defining *agile maturity*, e.g., [20], [28], [29]. Yet, so far, there is no standardized agile maturity model that is comparable with and as widely accepted as CMMI or ISO/IEC 15504. The results of these studies thus generate an incomplete picture, and numerous companies and project teams remain—or even become—skeptical and do not consider agile methods as the "Silver Bullet" [17], [22], [30], [31].

In 2011, West et al. [32] stated that modern software development evolved again. They coined the term *Water-Scrum-Fall* to reflect that software processes are in fact combinations of different traditional and agile methods and practices. Supporting this claim, Aitken and Ilango [33] state that *"there is nothing really incompatible"* with applying agility along with most traditional methods. A balanced combi- nation is needed, an assertion already stated by Boehm and Turner [34], who aimed to overcome the situation- specific shortcomings of agile and traditional development by defining five dimensions that describe a project environment. This helped to determine a balanced method and, eventually, to achieve a *hybrid sweet spot* [35]. These balanced combinations have become reality as several studies show [36]–[40]. Traditional and agile approaches coexist and form the majority of practically used *hybrid development methods* [8], [41], [42]. Specifically, Tell et al. [8] found hundreds of process variants which are composed of different agile and traditional methods and practices, concluding that, for instance, there is no "one and only" correct implementation of Scrum, and that agile methods and practices are not implemented by the book [43]. Noll and Beecham [42] hypothesize that the mindset of the company determines whether a project will adopt a purely agile approach.

---

[2] This research is based on the HELENA study (Hybrid dEveLopmENt Approaches in software systems development, online: https://helenastudy.wordpress.com), which is a large-scale international survey in which 75 researchers and practitioners from 25 countries participated. We give further details on the implementation of the HELENA study in Section 3.2.

[3] Royce is often considered the "inventor" of the Waterfall model, but, in fact, the term was coined later by Boehm

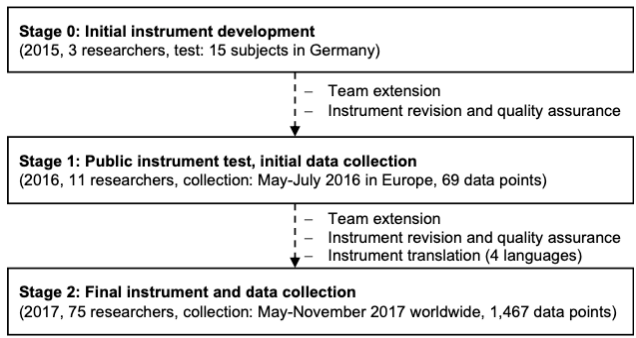

Fig. 1. Overview of the multi-staged data collection approach

Even though current literature provides an increasing amount of research focusing on agile methods and practices, it is still unclear what "agile" means. Is it the process model? Is it a mindset or a cultural question? Beck describes the driving concept of the process modeling behind Extreme Programming as: *"crank up all the knobs to 10 on the things I thought were essential and leave out everything else"[4]*, which points to a more behavioral perspective. Hence, we take the position that agile is a mindset, i.e., *the degree of agility is a personal perception of the managers and developers involved in a project* [27]. In this article, we seek to study this personal perception of agility through a data-driven characterization of reality, which captures details about individual processes reported by practitioners, as well as their perceived degree of agility [44]. We aim to overcome subjectivity as mentioned by Jalali et al. [27] and, therefore, to objectively identify those process elements that are more likely to be associated with agility. Our work does not aim to provide a precise definition of the term "agile software development". Instead, we provide information about individual perceptions and characterizations of agility which we quantify to determine their relation to software development processes.

## 3. RESEARCH DESIGN

We describe our research design by presenting the research questions in Section 3.1, the instrument development and data collection procedures in Section 3.2, the detailed description of the data analysis procedures in Section 3.3, and the validity threats in Section 3.4.

### 3.1 Research Objective and Research Questions

Research on agility and software processes is focused largely on frameworks, methods, and practices. However, *does that mean that methods and practices are the (only) building blocks for agility?* In this paper, we aim to better understand whether and how the degree of agility derives from the methods and practices used. We pose the following research questions:

**RQ 1:** What is the degree of agility in implementing typical project disciplines in software companies? Software development consists of a variety of activities grouped in project disciplines, such as project and quality management, architecture and design, implementation, and so forth [9]. Our first question addresses the degree of agility across these project disciplines. We aim to identify setups having either consistently low or consistently high degree of agility in their implementation of the project disciplines, which serve as input for studying the second research question.

**RQ 2:** Which methods and practices influence the degree of agility of implementing the project disciplines in software companies? The first research question provides an overview of the implementation of the different project disciplines, but it provides no details about the influence of specific methods and practices on the degree of agility. The second research question aims at statistically analyzing which methods and practices increase or decrease the degree of agility.

### 3.2 Instrument Development and Data Collection

We used the survey method [45] to collect our data. We designed an online questionnaire to collect data from practitioners about the processes they use in their projects. The *unit of analysis* was either a project or a software product.

#### 3.2.1 Instrument Development

We used a multi-staged approach to develop the survey instrument [41], which is illustrated in Fig. 1. Initially, three researchers developed the questionnaire and tested it with 15 German practitioners to evaluate its suitability. Based on the findings and feedback presented in [46], a team of 11 researchers from across Europe revised the questionnaire. The public test of the revised questionnaire, that included up to 25 questions, was conducted in 2016 in Europe and yielded 69 data points [39], which were analyzed and used to initiate the next stage. In Stage 2, the team was extended, with 75 researchers and practitioners from all over the world. The revision of the questionnaire for Stage 2 focused on the improvement of structure and scope, e.g., relevance and precision of the questions, value ranges for variables, and relevance of the topics included. Furthermore, the questionnaire, which was available at that stage in English only, was translated to German, Spanish, and Portuguese. Further details of the instrument are available in [44].

#### 3.2.2 Instrument Structure

The final questionnaire consisted of five parts: *Demographics* (group/question code D, 10 questions), *Process Use* (code PU, 13 questions), *Process Use and Standards* (code PS, 5 questions), *Experiences* (code EX, 2 questions), and *Closing* (code C, 8 questions). In total, the questionnaire consisted of up to 38 questions including the conditional questions that depend on previously given answers [44]. The questions on *Process Use* covered 24 methods and 36 practices that were derived from literature.

---
[4] Taken from an interview by informIT, March 23, 2001: http://www.informit.com/articles/article.aspx?p=20972, last access: February 6, 2019.

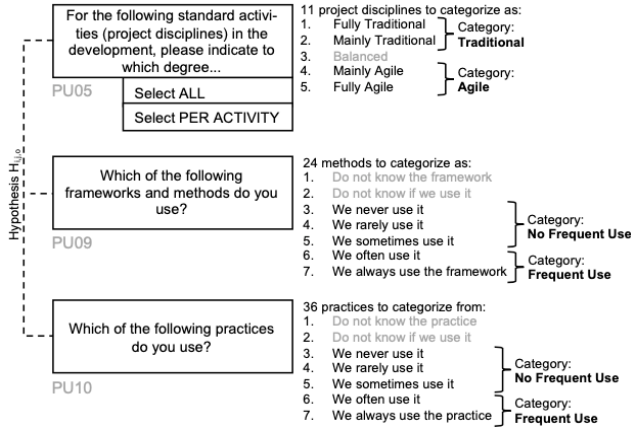

Fig. 2. Analysis model. The model shows the three questions (incl. question IDs), the value ranges and the linked hypotheses

### 3.2.3 Data Collection
The data collection period was May to November 2017 following a *convenience sampling strategy* [45]. The survey was promoted through personal contacts of the 75 participating researchers and practitioners, through posters at conferences, and through posts to mailing lists, social media channels (Twitter, Xing, LinkedIn), professional networks and websites (ResearchGate and personal home pages). In total, the survey yielded 1,467 responses (every response is a *data point* used for analyses). While the raw dataset and a basic characterization of the population can be obtained from [44], Section 4.1 provides a summary of the 556 data points selected for this study after cleaning and reducing the data (cf. Section 3.3.1).

### 3.3 Data Analysis Procedures
This section describes the analysis procedures in detail. We present the data cleaning and data reduction procedures, introduce our overall analysis model and provide detailed information on the procedures implemented to study our research questions.

### 3.3.1 Data Cleaning and Data Reduction
Due to the analysis model adopted as shown in Fig. 2, we had to discard some data points and aggregate data to support the different analyses. Hence, we execute the analyses with varying reported *n*-values.

To clean the dataset, we analyzed the data for *NA* and *-9* values captured by the online survey tool. While *NA* values indicate that participants did not provide information for optional questions, *-9* values indicate that participants skipped a question. The base dataset consisted of all data points that provided information for at least one of the project disciplines in PU05 (see Fig. 2), leading to a sample size of *n*=556. In the different analysis steps, we analyzed varied data sets (emerging from this base dataset), depending on (a) the project disciplines covered in the data points (analyses per project discipline for RQ1 and RQ2) and (b) the awareness that a specific method or practice is used (analyses per method or practice for RQ2).

### 3.3.2 Development of the Analysis Model
Figure 2 shows the analysis model which we developed to provide a framework for the analysis. It consists of three questions in the questionnaire. In the rest of the paper, we use short versions of the questions from Fig. 2 (together with the question ID to allow the mapping).

To analyze the degree of agility (RQ1) in the different project disciplines (PU05, following the SWEBOK categories [9]), we reduced the categories by merging and creating two new sets:

$$S_{traditional}(p) = S_{fully\_trad}(p) \cup S_{mainly\_trad}(p) \quad (1)$$

$$S_{agile}(p) = S_{fully\_agile}(p) \cup S_{mainly\_agile}(p) \quad (2)$$

S is the selection of the participants indicating whether they implement a specific project discipline p in a "more" traditional or agile fashion. All analyses use the two new selection sets $S_{agile}(p)$ and $S_{traditional}(p)$.

To answer RQ2, and in contrast to our previous studies [8], [41], we do not consider all methods (PU09) and practices (PU10) independent from their respective frequency of use (Fig. 2). Instead, we only use those methods and practices that have been used "often" or "always". We discard the "rarely" or "sometimes" used methods and practices to avoid noise introduced by exceptional cases. As we wish to analyze if the use of a method or practice has an influence on the degree of agility of the project disciplines (PU05), we test whether both data sets (consisting of all data points having reported on the degree of agility of the respective project discipline and on not or rarely using a specific method or practice and those data points that use this method often or always) belong to the same population.

We use the $\chi^2$ test to compare the distributions for each of the 24 methods and the 36 practices and their respective use per project discipline (11 disciplines). That is, we performed $(24 + 36) \times 11 = 660$ $\chi^2$ tests (one test per case where a case is a pair (method/practice, project discipline)). Each test analyzes the null hypothesis $H_{i,j,0}$, where $i \in \{1, \ldots, 60\}$ represents a specific method or practice and $j \in \{1, \ldots, 11\}$ represents one of the 11 project disciplines in a software-producing organization. In cases where we have few data points, we used Fisher's exact test, which can be applied instead of the $\chi^2$ test for small sample sizes. However, as the $\chi^2$ test—if applicable—is stronger than Fisher's exact test, we decided to use Fisher's exact test only if the $\chi^2$ test is inapplicable. Eventually, our null hypotheses have the form:

$$H_{i,j,0} : P(X_{i,j}) = P(X_{\bar{i},j}) \quad (3)$$

That is, we test if the distribution $X_{i,j}$ belongs to the same population as the distribution $X_{\bar{i},j}$. Hence, $X_{i,j}$ is the set of data points having reported on the degree of agility for project discipline $j$ and using method or practice $i$ often or always (Fig. 2). Likewise, $X_{\bar{i},j}$ is the set of data points having reported on the degree of agility for project discipline j and not or rarely using method or practice $i$.

Since we performed 660 tests analyzing the same hypothesis (with adjustments), we apply the Bonferroni correction to adjust the p-value accordingly. Compared to other corrections, e.g., Holm's step-down or Hochberg's step-up procedures, the Bonferroni correction is, according to Strass-burger and Bretz [47], the most pessimistic option

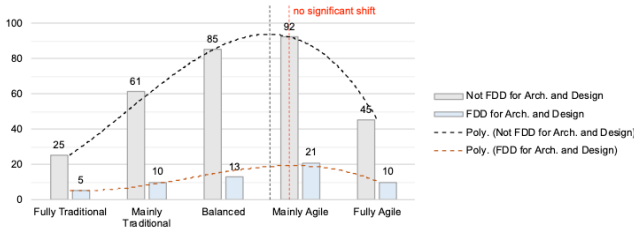

Fig. 3. Integrated histogram for the project discipline *Architecture and Design* (n=536) and for the method *Feature-Driven Development* (n=59) within the *Architecture and Design* discipline (note that, even though we illustrate this as a curve, the applied test works on categorical data)

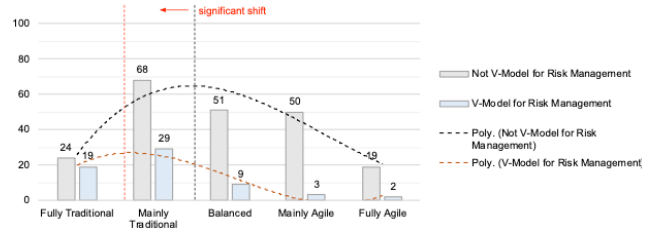

Fig. 4. Integrated histogram for the project discipline *Risk Management* (n=462) and for the method *V-shaped Process (V-Model)* (n=62) within the *Risk Management* discipline

leading to the smallest (less or equal) adjusted p-value, which is in our study $p_{corr} = \frac{0.05}{660} \approx 7.57 \times 10^{-5}$. We clearly highlight differences in the results if the correction is (not) applied.

*3.3.3 Specific Analysis Procedures for RQ 1*
Using our analysis model (Fig. 2), we first study the state of practice regarding the degree of agility of the project disciplines (PU05) in software producing organizations. We study, for instance, if project management is implemented in a more agile or in a more traditional fashion. For this, we calculate the distribution of the degree of agility per project discipline using all data points reporting the degree of agility in the respective project discipline. This results in 11 distributions of the degree of agility.

From our previous studies [8], [41] we know that approximately 75% of the participants use hybrid methods to run their projects. In this regard, in the second step, we study how many participants state that they implement *all* project disciplines in a "purely" traditional or agile manner. To get this specific distribution, we increase the strictness of the data point selection, i.e., analyze those data points only that claim to implement *all* project disciplines either fully or mainly agile or, respectively, fully or mainly traditional.

*3.3.4 Specific Analysis Procedures for RQ 2*
The different project disciplines are to a large extent implemented by a combination of different methods and practices [8], [41]. Therefore, an investigation of the impact on a project discipline's implementation needs to start with the question if a method or practice is used in that project discipline. Our data does not provide such a direct mapping, as an assessment of these details would have unacceptably increased the length of the questionnaire. Yet, we can derive such information from the data using two assumptions:

**Assumption 1:** A method or practice that has no influence on the degree of agility of the project discipline of interest will not change the distribution of cases with different degrees of agility compared to the project discipline in general.

**Assumption 2:** A method or practice that has an effect on agility, but is not used in the context of a project discipline is not likely to correlate with a shift in the distribution of degrees of agility of that project discipline.

Figure 3 illustrates *Assumption 1*. The figure provides two integrated histograms, which show (a) the distribution of the degree of agility in the project discipline *Architecture and Design* among all data points that use *Feature-Driven Development* (FDD), and (b) the distribution of the degree of agility in the discipline *Architecture and Design* among all data points that do not use FDD to embody this discipline. The two trend-lines in Fig. 3 indicate that FDD appears to be relevant in this discipline, yet, FDD seems to have no impact on the degree of agility, as the distribution shows the tendency, revealing only marginal differences.

Figure 4 illustrates *Assumption 2*. The figure shows the distribution of the degree of agility for the project discipline *Risk Management* among all data points, and the distribution of the degree of agility in the discipline *Risk Management* among all data points that use the *V-shaped Process (V- Model)* to embody this discipline. The integrated histogram indicates a shift in the distribution, i.e., the *V-Model* is a relevant method in *Risk Management*, but its use causes a shift towards a lower degree of agility. However, the second assumption has to be taken with care, as a spurious correlation might exist due to confounding factors.

Since a visual inspection is not sufficient, we use statistical tests to compare the distributions of the degree of agility for a project discipline between the cases that use and that do not use a practice or method (Eq. 3). For this, we calculate the frequencies of different degrees of agility per project discipline of the data points using a specific method or practice, e.g., *Scrum* or *Pair Programming*. For the examples in Fig. 3 and Fig. 4, the χ2 test results are:

- Project discipline *Architecture and Design* using *Feature-Driven Development*: *p*-value=0.827 > $7.57 \times 10^{-5}$. That is, the use of *Feature-Driven Development* does not significantly change the degree of agility of the discipline *Architecture and Design*, as there is no significant difference in the distributions.
- Project discipline *Risk Management* using *V-shaped Process (V-Model)*: *p*-value=$1.28 \times 10^{-5} < 7.57 \times 10^{-5}$. That is, the use of the *V-Model* significantly changes the degree of agility of the discipline *Risk Management*, as there is a significant difference in the distributions.

These results support the impressions taken from the visual inspection of Fig. 3 and Fig. 4. In the case of significant results, i.e., a *p*-value< $7.57 \times 10^{-5}$ for the scenario in Fig. 4, the test indicates only that a specific method or practice leads to a shift. However, we cannot characterize this shift, i.e., whether this is a shift towards agile or traditional. We only know that a specific method or practice is likely to have an impact on the degree of agility. That is, in a first step, we identify the *influencing methods and practices* without quantifying the actual influence.

To analyze and quantify the influence (the shifts) in more detail, we compared the median degrees of agility for the distribution limited to the rare use, if at all, of the respective method or practice with the distribution limited to the use of the respective method or practice. For example, for *Risk Management* (Fig. 4), the median degree of agility is 3 for data points not working with the *V-Model* and 2 for those working with the *V-Model*, thus indicating a tendency towards a more traditional development approach. In other words, using the *V-Model* makes *Risk Management* in a project "more traditional". In several cases the median does not change despite a significant difference in the distribution. In these cases, we assume that the extent of the difference is negligible.

### 3.4 Threats to Validity
We discuss the threats to validity following the classification presented in Wohlin et al. [48].

*Construct validity:* The dataset analyzed in this paper emerged from a survey. One of the main threats of survey-based research is the risk of misunderstood questions leading to incomplete or wrong responses. To mitigate this risk, several researchers were involved in designing and revising the questionnaire, including pre-tests, internal and external reviews as described in Section 3.2.1. In addition, native speakers from the team translated the English version of the questionnaire into German, Spanish, and Portuguese to reduce the risk of misunderstandings due to language issues. We distributed the questionnaire using the convenience sampling strategy [45] as described in Section 3.2.3. This introduced the risk of losing control in terms of sampling, response rate and so forth. To ensure that the participants still represent our target population, we implemented rigorous data pre-processing including consistency checks. Analyzing free-text questions in the questionnaire led to reasonable results [41]. Hence, we are confident that this threat can be considered mitigated.

Software engineering is challenged with terminological confusion [49]. Study participants will as a result have somewhat varying understandings of *agile* software development, however, the practices included in this research are widely used (as demonstrated by the survey responses), they support agile principles as identified in the Agile Manifesto [1], and as such support an examination of hybrid software engineering. We mitigated that threat by analyzing a large population and quantifying the perceptions using an explicit metric (Likert scale, Section 3.3.2). However, even the aggregated results could be biased and, therefore, require independent research for confirmation.

*Internal Validity:* Threats to internal validity were potentially introduced while preparing and cleaning the data. Also, the selection of the statistical tests can threaten internal validity. To mitigate these risks, all analysis steps have been performed by at least two researchers and reviewed by at least two other researchers not involved in the actual analyses. The remaining researchers in the team were asked to provide quality assurance. Implementing these rigorous review processes, we are confident that our research method is reliable and can be reproduced. Another threat to internal validity might have been introduced by the dataset itself. Analyzing the overall degree of agility in the dataset, we find a general tendency towards agile that is potentially caused by the data collection procedure. Hence, every sub-set used in the data analysis can be biased towards this direction. We discuss potential effects in Section 5.

The implementation of the analysis model (Section 3.3.2) might introduce another threat: we include data points in multiple analyses, e.g., when analyzing the influence in dependence of the industry target domain. If a participant selected multiple industry target domains, this data point was considered for each analysis. Consequently, the applied statistical tests are executed on a potentially biased dataset (base population) and, thus, the impact of specific methods and practices could have been underestimated. As the statistical tests analyze differences in the distributions of specific datasets, this bias may have influenced the results. However, as we do not claim generalizability, we discuss potential effects in Section 5.

*Conclusion Validity:* Since we perform 660 tests for the same hypothesis and using the same dataset (Section 3.3.2), we applied the Bonferroni correction leading to an adjusted significance level of $p_{corr} \approx 7.57 \times 10^{-5}$ (we highlight results still significant using the adjusted p-value instead of the normal p-value of 0.05). To analyze the effect of the significant results, we compared the median values of the respective distributions, assuming that the effect size of the difference is negligible for same median values. In the other cases, the shift in the median indicates the direction of the effect of a specific method or practice towards agile or towards traditional. Nonetheless, the identified significant results have to be confirmed in future studies.

*External Validity:* Our data analysis is based on a dataset comprised of 1,467 data points. Nevertheless, due to the data collection strategy (Section 3.2.3) and its impact on the population, we cannot claim generalizability. In some cases, we find statistically significant differences between two sub-sets. However, as we lack sufficient data to provide a solid conclusion, we consider these observations as candidates for which further research is necessary to confirm the results.

### 4. RESULTS
This section is structured according to the two research questions as introduced in Section 3.1. The interpretation and discussion of the results are provided in Section 5.

### 4.1 Demographics
Before we present the study results, in this section, we briefly describe the study population. As described in Section 3.2.3, the survey yielded 1,467 data points of which we selected 556 for this study (Section 3.3.1). Among the 556 participants, 555 provided information about their company size: 133 (23.92%) work in micro or small companies with less than 51 employees, 137 (24.64%) work in medium-sized companies (≤250 employees), 156 (28.06%) work in large companies with up to 2.500 employees, and 129 (23.20%) work in very large companies with more than 2.500 employees. In total, 351 out of 556 participants (63.13%) state that they are involved in distributed development (regionally, nationally, and globally). Finally, 336 (60.43%) of the participants state that they work on very large software projects, i.e., projects that have a staffing level of more than one person year. Another 189 participants (33.99%) are involved in medium to large projects with staffing levels of two person months to one person year.

| Role / Experience | <1 year | 1-2 years | 3-5 years | 6-10 years | >10 years |
|---|---|---|---|---|---|
| Analyst/Requirements Engineer | 1 | 3 | 3 | 4 | 18 |
| Architect | 0 | 2 | 3 | 8 | 35 |
| C-level Management | 0 | 1 | 2 | 4 | 31 |
| Developer | 5 | 13 | 37 | 38 | 49 |
| Product Manager/Owner | 1 | 2 | 5 | 11 | 31 |
| Project/Team Manager | 2 | 3 | 10 | 22 | 76 |
| Quality Manager | 0 | 0 | 1 | 2 | 26 |
| Scrum Master/Agile Coach | 3 | 3 | 3 | 2 | 28 |
| Tester | 1 | 0 | 2 | 1 | 3 |
| Trainer | 0 | 0 | 1 | 0 | 5 |
| Other | 2 | 2 | 9 | 6 | 36 |

Fig. 5. Participant roles and experience (alphabetically sorted)

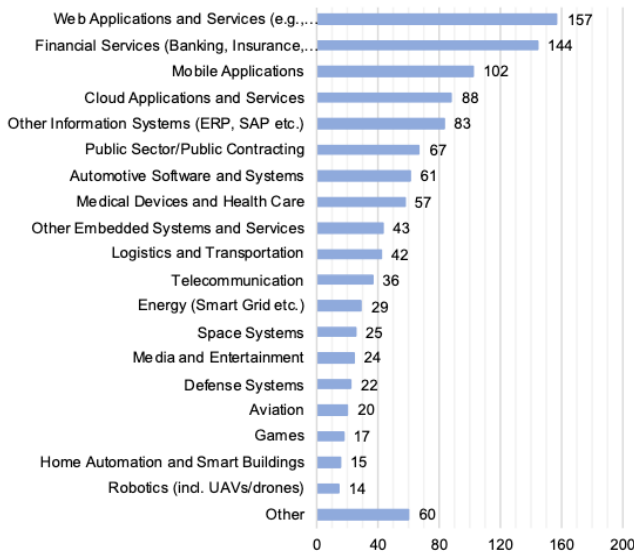

Fig. 6. Overview of the application domains of the project/products

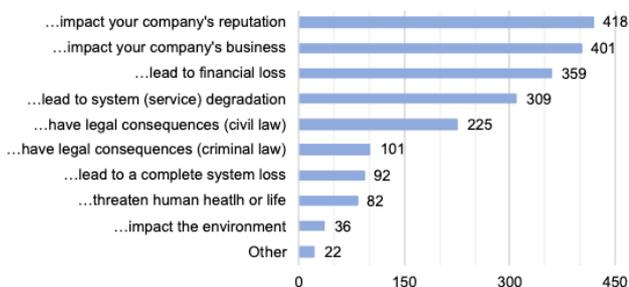

Fig. 7. Overview of the different criticality levels of the project/products

The study participants have different roles reflecting a good bandwidth of skills combined with a high level of experience (Fig. 5). Developers (142; 25.54%) and project/team managers (113; 20.32%) are the largest participant groups, and a majority of the participants (338; 60.79%) has more than 10 years of professional experience.

The survey's unit of analysis was a project or product in which the participants are involved. The projects/products address different application domains, which are summarized[5] in Fig. 6. The figure shows that *Web Applications and Services* (157; 28.24%), *Financial Services* (144; 25.90%), and *Mobile Applications* (102; 18.35%) are the most frequently mentioned application domains. Also, 60 participants (10.79%) mentioned they work in other do- mains such as Agriculture, Geo-Information Systems or CAD/Electronic Design. Furthermore, participants were asked about the importance of software and systems development, i.e., the criticality of software to their company. Figure 7 shows that an impact on reputation and business are the most frequently mentioned risks. Yet, 82 participants (14.75%) state that issues of the software could impact human health or life and another 36 (6.47%) see risks to the environment, which underlines the critical role software has gained today.

### 4.2 RQ 1: Degree of Agility of Project Disciplines

To answer the first research question, we evaluate the degree of agility per project discipline as reported by the study participants.

Figure 8 provides an integrated perspective by presenting one histogram for each project discipline. The histograms show that there is no general trend evident towards *purely* agile or traditional development. Instead, a general tendency towards a hybrid development approach can be noted (as already observed in [8], [39], [41]). The disciplines *Project Management*, *Change Management*, *Requirements Engineering*, *Architecture and Design*, *Implementation/Coding*, *Integration and Testing*, and *Maintenance* show a tendency towards a more agile implementation.

Yet, Fig. 8 does not provide information about the consistent implementation of the respective project disciplines, i.e., if a project is undertaken in a consistently traditional or agile way. In Fig. 9, we provide an aggregated perspective on the consistent use of agile or traditional methods and practices. We selected all data points for which we found a consistent selection of the degree of agility, i.e., for all project disciplines, participants selected the same category, e.g., *Fully Agile* or *Mainly Agile*. The figure shows that few participants implement their projects consistently traditional (7) or consistently agile (9). Another 14 participants claim to consistently implement their projects in a balanced manner. Figure 9 shows that 46 participants implement their projects in an agile style and, of these, nine participants do it "fully agile", 21 "mainly agile", and another 16 participants implement their projects either fully or mainly agile, i.e., at least mainly agile. In total, from the 556 analyzed data points, 83 (14.93%) claim to consistently implement the different project disciplines fully or mainly agile or traditional, or in a balanced fashion. With some 15% of the studied population, the "pure doctrine" accounts for a small share only.

---

[5] Please note that the participants could choose multiple options for the application domains (Fig. 6) and the criticality levels (Fig. 7).

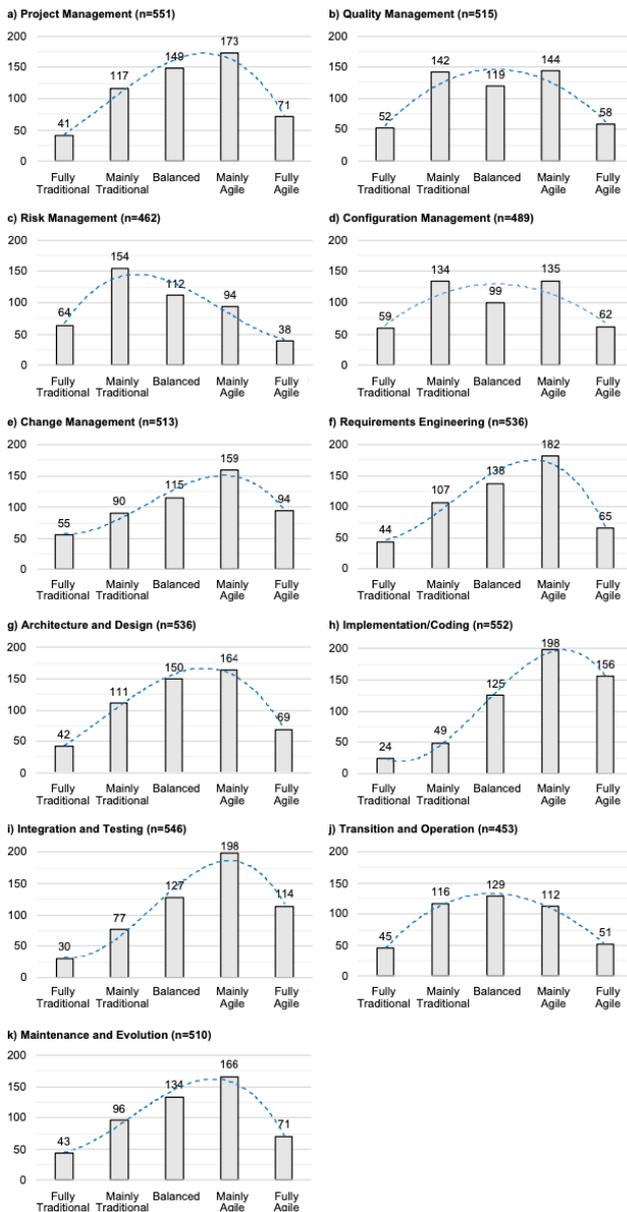

Fig. 8. Number of data points in dependence of the degree of agility per project discipline (n=556)

**Finding 1a:** Most project disciplines show a clear trend towards an agile implementation. Exceptions are the three project disciplines *Quality Management*, *Configuration Management* and *Transition and Operation*, which are balanced, as well as the discipline *Risk Management* that has a trend towards traditional development.

**Finding 1b:** A small share of approx. 15% of the participants implement all project disciplines consistently, either agile, traditional, or balanced.

### 4.3 RQ 2: Influence on the Degree of Agility

To answer the second research question, we compare the distributions of the degrees of agility in relation to the frequency of use of a specific method or practice. For this, we first identify those methods and practices causing a shift, before we characterize the found shifts.

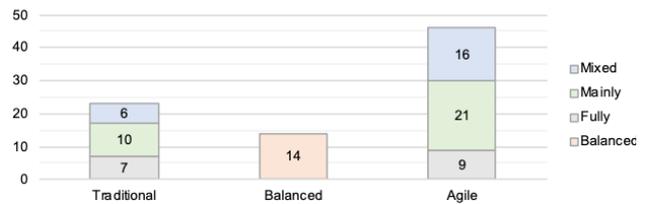

Fig. 9. Aggregated degree of agility for participants implementing all project disciplines consistently (n=83)

*4.3.1 Identifying Influencing Methods and Practices*
Figure 10 presents the results of the $\chi^2$ tests and, if there were too few data points, the results of the Fisher's exact tests (as described in Section 3.3.2).

Figure 10 shows 146 cases where the frequent use of a specific method or practice changes the distribution of the degrees of agility (blue numbers). For example, using *Waterfall* or *Scrum* changes the distribution for all project disciplines. Yet, other methods and practices seem to have no influence, e.g., *Nexus*, *DSDM*, and *RUP*. Furthermore, Fig. 10 shows that practices seem to have a greater influence on the degree of agility than the methods, as they are more likely to result in a significant shift. Compared to the methods, we observe more significant differences in the distributions of the degrees of agility, i.e., $p < 7.57 \times 10^{-5}$. Specifically, from the 24 methods, two (Scrum and Waterfall) have an impact on all disciplines, and from the 36 practices, one (Formal Specification) has an impact on all disciplines. These numbers change to five methods and 15 practices influencing all project disciplines at a significance level of $p < 0.05$ (black numbers in Fig. 10). Furthermore, the disciplines are impacted differently. For instance, 20 out of 60 methods and practices have an impact on the discipline *Implementation and Coding*, followed by *Integration and Testing* (18), and *Architecture and Design* (15).

**Finding 2:** Compared to the selection of methods used, the selection of practices used has a stronger effect on the degree of agility of a given discipline.

**Finding 3a:** Few methods and practices affect the degree of agility of all project disciplines. These are marked in Fig. 10 in green (3; $p < 7.57 \times 10^{-5}$) and yellow (17; $p < 0.05$).

**Finding 3b:** The project discipline *Implementation and Coding* is most likely to be affected in its degree of agility by changes in the selected methods and practices, followed by *Integration and Testing*, and *Architecture and Design*. The project discipline least likely to be affected is *Risk Management*.

*4.3.2 Quantifying Influencing Methods and Practices*
While Fig. 10 summarizes significant differences in the degree of agility depending on method and practice selection, the figure does not show *how* these differences are manifested. For this, as described in Section 3.3.4, we study the median values of the degrees of agility per project discipline as reported by the participants. This comparison includes the frequency of use of a specific method or

| | | Project Management | Quality Management | Risk Management | Configuration Management | Change Management | Requirements Analysis/ Engineering | Architecture and Design | Implementation and Coding | Integration and Testing | Transition and Operation | Maintenance and Evoution |
|---|---|---|---|---|---|---|---|---|---|---|---|---|
| Methods | *Waterfall* | 0.00E+00 | 8.00E-13 | 1.42E-08 | 1.92E-05 | 5.71E-13 | 1.44E-13 | 2.87E-13 | 3.24E-16 | 1.74E-15 | 3.92E-08 | 1.85E-12 |
| | Crystal | | | | | | | | | | | |
| | *DevOps* | 1.67E-08 | 4.71E-06 | 5.90E-05 | 5.65E-10 | 2.51E-07 | 1.44E-04 | 4.61E-08 | 8.75E-08 | 2.51E-08 | 3.57E-10 | 1.06E-04 |
| | Domain-Driven Design | | | | | | | 2.23E-02 | 3.53E-02 | | | 3.62E-02 |
| | DSDM | | | | | | | | | | | |
| | *Extreme Programming* | 5.40E-05 | 1.66E-05 | 1.19E-03 | 4.83E-05 | 2.98E-05 | 5.76E-03 | 1.08E-02 | 1.93E-06 | 8.68E-10 | 1.26E-08 | 2.74E-05 |
| | Feature-driven Development | 3.45E-02 | | | | | | | | | | |
| | Iterative Development | 4.34E-03 | 2.66E-02 | 3.37E-02 | 1.69E-03 | 3.18E-04 | | | 1.84E-05 | 2.62E-04 | 1.96E-02 | 8.71E-03 |
| | Kanban | 2.49E-02 | 4.66E-02 | 1.19E-02 | | | 4.55E-02 | | 9.74E-03 | 1.03E-02 | 5.17E-04 | 5.95E-03 |
| | Large-scale Scrumm (LeSS) | | 3.41E-02 | | | | | | | 9.82E-03 | 4.00E-02 | |
| | Lean Software Development | 1.36E-02 | | 3.84E-03 | 6.87E-03 | 1.27E-02 | 1.30E-02 | 9.56E-05 | 1.03E-03 | 1.59E-04 | 8.48E-05 | 5.19E-03 |
| | Model-Driven Architecture (MDA) | | | | | | | | 3.10E-02 | | | |
| | Nexus | | | | | | | | | | | |
| | Personal Software Process (PSP) | | | | | | | | | | | |
| | Phase / Stage-gate model | 9.78E-04 | 9.85E-04 | 9.85E-03 | | | | | | 2.53E-02 | 3.81E-02 | 9.19E-03 |
| | PRINCE2 | | | | | | | | | | | |
| | Rational Unified Process (RUP) | | | | | | | | | | | |
| | Scaled Agile Framework (SAFe) | | | | 9.91E-03 | | | | | | | |
| | *Scrum* | 3.10E-10 | 4.90E-05 | 5.23E-05 | 4.73E-08 | 3.73E-08 | 2.85E-08 | 4.80E-07 | 2.97E-14 | 8.03E-12 | 3.52E-07 | 6.36E-07 |
| | ScrumBan | 5.13E-03 | 4.01E-02 | 4.51E-02 | 1.08E-02 | | 9.20E-04 | 3.10E-02 | 4.32E-02 | 1.43E-03 | | |
| | Spiral Model | | | | | | | 1.76E-02 | 1.11E-02 | | | |
| | SSADM | 4.69E-02 | 2.69E-02 | | | | 2.01E-03 | 4.10E-02 | 2.48E-03 | 2.71E-03 | 2.68E-02 | |
| | Team Software Process (TSP) | | | | | | 1.77E-02 | | 1.43E-02 | | | |
| | *V-Shaped Process (V-Model)* | 5.50E-05 | 6.03E-09 | 1.28E-05 | 1.44E-04 | 3.58E-04 | 1.66E-05 | 1.01E-05 | 1.22E-04 | 4.54E-03 | 4.00E-04 | 1.36E-05 |
| Practices | Architecture Specifications | 1.09E-02 | 3.24E-02 | 3.28E-02 | 2.22E-02 | 7.75E-03 | 2.30E-02 | 6.83E-04 | | | 1.03E-02 | 3.61E-02 |
| | Automated Code Generation | | | | | | | | | | | |
| | Automated Theorem Proving | | | | | | | | 1.75E-02 | | | |
| | Automated Unit Testing | 4.09E-04 | 2.87E-03 | | 3.12E-03 | | | 2.74E-02 | 2.83E-06 | 2.18E-06 | 7.31E-03 | 1.10E-02 |
| | *Backlog Management* | 7.13E-10 | 3.28E-06 | 7.84E-05 | 2.00E-04 | 1.87E-07 | 4.52E-07 | 5.43E-06 | 8.26E-12 | 8.91E-09 | 1.00E-05 | 1.41E-07 |
| | Burn-Down Charts | | | | | | | | 1.59E-03 | | | |
| | Code Reviews | 1.47E-03 | 2.15E-02 | | | | 7.17E-03 | 8.74E-03 | 6.06E-04 | 4.31E-04 | | 8.13E-03 |
| | Coding Standards | | | 2.69E-02 | | | | 8.36E-03 | | | | 4.16E-02 |
| | *Collective Code Ownership* | 8.35E-05 | 2.24E-04 | 2.07E-03 | 1.26E-06 | 9.39E-08 | 5.29E-04 | 5.44E-06 | 6.58E-07 | 1.15E-04 | 2.41E-03 | 5.65E-05 |
| | *Continuous Deployment* | 2.75E-04 | 5.35E-05 | 1.32E-04 | 6.23E-06 | 1.10E-05 | 6.31E-05 | 1.71E-03 | 9.09E-05 | 5.34E-09 | 2.24E-09 | 1.66E-04 |
| | *Continuous Integration* | 5.36E-06 | 6.97E-06 | 1.77E-04 | 2.47E-05 | 8.70E-06 | 1.52E-07 | 1.63E-06 | 1.05E-07 | 6.26E-11 | 2.20E-05 | 6.86E-06 |
| | *Daily Standup* | 8.02E-07 | 7.42E-04 | 2.72E-02 | 2.13E-03 | 9.96E-06 | 1.86E-06 | 3.48E-05 | 6.49E-11 | 2.41E-09 | 4.68E-05 | 3.37E-04 |
| | *Def. of Ready/Done* | 1.17E-05 | 2.01E-04 | 2.17E-04 | 7.57E-06 | 1.27E-06 | 1.22E-05 | 1.28E-04 | 3.67E-09 | 2.00E-07 | 4.38E-07 | 2.50E-06 |
| | Design Reviews | | | 3.59E-03 | | | | | 3.14E-02 | | | |
| | Destructive Testing | | | | | | | | 1.16E-02 | | | |
| | *Detailed Designs* | 6.22E-03 | 3.80E-04 | 2.06E-03 | 5.71E-03 | 4.30E-03 | 3.78E-04 | 1.67E-07 | 1.23E-07 | 2.07E-05 | 4.26E-02 | 1.17E-03 |
| | Limit Work-in-Progress | 1.59E-02 | 1.17E-02 | | | | 1.66E-02 | 5.11E-03 | 1.08E-02 | 3.38E-03 | 1.45E-03 | 6.92E-04 |
| | End-to-End (System) Testing | | | | | | | | | | | |
| | Expert/Team based estimation | 1.77E-02 | 2.12E-03 | 6.53E-03 | 3.96E-05 | 2.80E-03 | 2.02E-03 | 5.41E-04 | 1.66E-05 | 2.81E-04 | 1.18E-03 | 3.34E-02 |
| | Formal estimation | | 1.49E-02 | | | 2.44E-02 | 3.88E-03 | 4.49E-05 | 2.70E-04 | 8.89E-03 | 1.44E-02 | |
| | *Formal Specification* | 8.68E-07 | 9.36E-07 | 4.34E-05 | 2.57E-06 | 7.12E-08 | 8.41E-09 | 1.40E-09 | 4.42E-07 | 7.47E-07 | 6.91E-05 | 1.69E-08 |
| | Iteration Planning | 1.27E-02 | | | | | 4.53E-02 | 3.30E-02 | 6.19E-05 | 1.08E-02 | 3.78E-02 | |
| | *Iteration / Sprint Reviews* | 4.31E-09 | 2.06E-04 | 1.85E-04 | 4.82E-04 | 4.49E-05 | 4.95E-08 | 8.32E-07 | 1.13E-10 | 3.16E-07 | 8.82E-06 | 3.80E-06 |
| | Model Checking | | | | | | 2.86E-02 | 4.39E-02 | | | | |
| | On-Site Customer | 3.59E-02 | | | | | 1.15E-02 | | 2.63E-02 | | | |
| | Pair Programming | 1.42E-02 | | | 2.95E-03 | 1.51E-02 | 2.11E-02 | | 5.65E-04 | 1.01E-02 | | 4.01E-03 |
| | Prototyping | | | | | 1.19E-02 | 8.35E-03 | | | | | |
| | *Refactoring* | 9.36E-05 | 2.12E-05 | 5.91E-03 | 1.97E-06 | 8.01E-06 | 7.56E-06 | 1.30E-07 | 1.55E-09 | 1.87E-12 | 5.19E-05 | 1.80E-04 |
| | Release planning | | | | | | | | | | | |
| | *Retrospectives* | 3.54E-06 | 2.36E-04 | 3.05E-02 | 2.08E-04 | 2.11E-06 | 1.19E-06 | 2.92E-05 | 3.92E-12 | 1.37E-09 | 4.89E-04 | 4.59E-05 |
| | Scrum of Scrums | | | 1.33E-02 | 4.79E-02 | | | | | 5.60E-03 | 1.27E-02 | |
| | Security Testing | | | | | | | | | | | |
| | *Test-driven Development* | 1.06E-02 | 4.56E-03 | 5.93E-03 | 5.94E-04 | 1.64E-02 | 3.77E-02 | 4.66E-04 | 2.03E-02 | 5.11E-06 | 7.49E-07 | 2.62E-04 |
| | *User Stories* | 5.17E-04 | 4.39E-04 | 1.04E-03 | 3.21E-03 | 1.03E-03 | 4.15E-07 | 2.07E-04 | 2.71E-05 | 2.43E-05 | 2.35E-02 | 1.89E-03 |
| | *Velocity-based Planning* | 1.16E-04 | 6.35E-06 | 1.26E-05 | 1.63E-05 | 8.37E-04 | 1.05E-06 | 4.03E-06 | 7.84E-10 | 3.14E-06 | 2.99E-04 | 1.56E-03 |
| | Use Case Modeling | 1.91E-02 | | | | | 1.71E-02 | 4.09E-02 | | | | |

Fig. 10. Summary of the significant results with $p < 0.05$ (black) and $p < p_{corr} = 7.57 \times 10^{-5}$ (blue) of the $\chi^2$ tests. A green background indicates methods and practices on all project disciplines at a significance level of $p < p_{corr} = 7.57 \times 10^{-5}$ and, likewise, the yellow background indicates a significance level of $p < 0.05$. Gray-colored cells indicate the use of Fisher's exact test as described in Section 3.3.2

practice and targets participants that use or do not use a method or practice. As differences occur at different levels of significance, and even non-significant trends can be found, we integrated these information pieces into a color-coded and flagged representation. Figure 11 provides a guideline on how to read the resulting charts.

The color code shows the "starting state" of those participants that use a specific method or practice, i.e., what the median degree of agility is for this group. The flag shows the "trend" when comparing this group to the other group that does not use this method or practice. While the flags tr and ag stand for traditional and agile (see also Fig. 4), the flag n requires an extra explanation. This flag indicates that there is a significant difference. However, this difference does not indicate a specific trend as also explained by Fig. 3. In the following section, we first discuss the significant results found in Section 4.3.1, before we present the findings regarding observed trends that are not significant.

4.3.2.1 Corrected Significance: Figure 12 summarizes the trends and the shifts for the significant differences (using $p_{corr}$). The figure also shows those methods and practices for which no significant trend could be found at all (gray-italic items). In total, of the 660 cases, 146 (22.12%) show a significant difference, i.e., a potential shift in the degree of agility. Figure 12 shows that most of the trends identified cause an agile shift (88 out of 146, 60.27%). That is, applying a specific method or practice increases the degree of agility. This effect can be especially observed for cases that already have an agile starting state. Likewise, if a shift towards traditional occurs, the starting state is either *balanced* or *traditional*. Yet, a few methods only show a shift towards *traditional*, such as *Waterfall* and *Formal Specifications*.

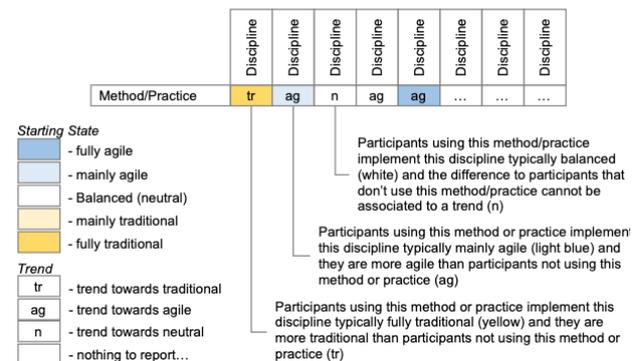

Fig. 11. Reading guideline for the trend analysis in Figures 12 – 14

Fig. 12. Significant shifts ($p < 7.57 \times 10^{-5}$) in the degree of agility

Fig. 13. Significant shifts ($p < 0.05$) in the degree of agility.

> **Finding 4:** Most practices are associated with an increase in the degree of agility.

Figure 12 also shows that there is no "radical" shift. That is, there is no case that, for instance, is usually *traditional* and shifts to *agile* when using a specific method or practice. This is in line with Noll and Beecham [42], who found method combinations tend to "stay in their class", e.g., findings are confirmatory, that projects that mainly use *agile methods* are more likely to use *agile practices*. Yet, no single practice determines whether a project is agile or not.

> **Finding 5a:** Methods and practices have a stable influence towards either a high or low degree of agility, which does not change with the project discipline.
>
> **Finding 5b:** No method or practice determines whether a project is traditional or agile, i.e., any method or practice can be found in traditional and agile development.

Figure 12 also shows that 36 methods and practices, seemingly, have no significant impact at all (using $p_{corr}$). Furthermore, for 29 cases, a significant difference could be found, but no tendency towards *agile* or *traditional* could be found (all entries in Figure 12 with entry "n"). These 29 cases represent methods and practices that show no tendency in the context of a specific project discipline only, e.g., *Scrum* (*Quality Management* and *Architecture and Design*). This means that using these methods or practices makes no difference, this method or practice is not impacting agility in general or in specific contexts.

> **Finding 6:** Some methods and practices are "neutral". That is, they are not associated with a changing perception of the degree of agility. These are marked gray in Fig. 12.

4.3.2.2 Non-Corrected Significance: Figure 13 summarizes the trends and shifts for the differences at a significance level of $p < 0.05$ that do not satisfy the adjusted *p*-value $p_{corr}$. In total, of the 660 studied cases, 209 (31.67%) show a difference. Figure 13 shows that most of the trends identified cause an agile shift (89 out of 209, 42.58%) and

| | Project Management | Quality Management | Risk Management | Configuration Management | Change Management | Requirements Analysis/ Engineering | Architecture and Design | Implementation and Coding | Integration and Testing | Transition and Operation | Maintenance and Evolution |
|---|---|---|---|---|---|---|---|---|---|---|---|
| *Waterfall* | | | | | | | | | | | |
| Crystal | ag | ag | ag | | | ag | | ag | | | |
| *DevOps* | | | | | | | | | | | |
| Domain-Driven Design | ag | | | | | ag | | | | ag | |
| DSDM | ag | | | ag | | ag | | | | ag | |
| *Extreme Programming* | | | | | | | | | | | |
| Feature-driven Development | | | | | | ag | ag | | | | |
| Iterative Development | | | | | | ag | | | | | |
| Kanban | | | | | | | ag | ag | | | |
| Large-scale Scrumm (LeSS) | ag | | | | | ag | ag | ag | | | |
| Lean Software Development | | ag | | | | | | | | | |
| Model-Driven Architecture (MDA) | | | tr | tr | | | | | | | |
| Nexus | ag | ag | ag | | | | | | | | |
| Personal Software Process (PSP) | | tr | | | | tr | | | tr | | |
| Phase / Stage-gate model | | | | tr | tr | | | | tr | | |
| PRINCE2 | tr | | | | | | | | tr | tr | |
| Rational Unified Process (RUP) | | | tr | | | | | | tr | | |
| Scaled Agile Framework (SAFe) | ag | ag | | | | ag | ag | ag | | | |
| *Scrum* | | | | | | | | | | | |
| ScrumBan | | | | | | ag | | | | ag | ag |
| Spiral Model | | | tr | | tr | | | tr | | | |
| SSADM | | | | tr | | | | | | | |
| Team Software Process (TSP) | tr | tr | tr | | | | | | tr | | |
| *V-Shaped Process (V-Model)* | | | | | | | | | | | |
| *Architecture Specifications* | | | | | | | | | | | |
| Automated Code Generation | | tr | | | tr | | | | | | |
| Automated Theorem Proving | ag | | ag | ag | | | | | tr | | tr |
| Automated Unit Testing | | ag | | | ag | ag | | | | | |
| *Backlog Management* | | | | | | | | | | | |
| Burn-Down Charts | | | | | | | | ag | | | |
| Code Reviews | | ag | | | ag | | | | | | |
| Coding Standards | | | | | | | | ag | | | |
| *Collective Code Ownership* | | | | | | | | | | | |
| *Continuous Deployment* | | | | | | | | | | | |
| *Continuous Integration* | | | | | | | | | | | |
| *Daily Standup* | | | | | | | | | | | |
| *Def. of Ready/Done* | | | | | | | | | | | |
| Design Reviews | | | | | | tr | | | | | |
| Destructive Testing | | | | | | | | | tr | | |
| *Detailed Designs* | | | | | | | | | | | |
| Limit Work-in-Progress | | | | | | ag | | | | | |
| End-to-End (System) Testing | | | | | | ag | | | | | |
| *Expert/Team based estimation* | | | | | | | | | | | |
| Formal estimation | | | | tr | tr | | | | | | |
| *Formal Specification* | | | | | | | | | | | |
| Iteration Planning | | | | | | ag | | | | | ag |
| *Iteration / Sprint Reviews* | | | | | | | | | | | |
| Model Checking | | tr | tr | tr | | | | | tr | | |
| On-Site Customer | | | | | | ag | | | | | ag |
| Pair Programming | | | | | | | | | | | ag |
| *Prototyping* | | | | | | | | | | | |
| *Refactoring* | | | | | | | | | | | |
| Release planning | | | | | | ag | | | | | |
| *Retrospectives* | | | | | | | | | | | |
| Scrum of Scrums | ag | | | | | ag | ag | | | | ag |
| Security Testing | | | | | | | | tr | | | |
| *Test-driven Development* | | | | | | | | | | | |
| *User Stories* | | | | | | | | | | | |
| *Velocity-based Planning* | | | | | | | | | | | |
| Use Case Modeling | | | | tr | tr | | | | | | |

Legend:
- fully traditional / mainly traditional
- balanced
- fully agile / mainly agile

Fig. 14. Shifts in the degree of agility (without significant shifts)

that 84 out of 209 (40.19%) do not show a notable trend. It can also be observed that there is still no "radical" shift. That is, even at this more coarse-grained level of abstraction (compared to Fig. 12), there is no case that started in *traditional* and shifts to *agile*.

> **Finding 7:** Using the weaker significance level of $p < 0.05$, we find many more tendencies compared to Finding 4, but we still find no radical shift.

Figure 13 shows only one special case: *Design Reviews*. In the project discipline *Risk Management*, this practice is considered *fully traditional* and shows a trend towards *traditional*. Yet, the same practice is considered *mainly agile* in the context of the project discipline *Implementation and Coding*, but with no explicit tendency.

> **Finding 8:** Combining all results at a significance level of $p < 0.05$, we find 10 methods and practices that have no impact at all. This strengthens Findings 5a and 5b.

4.3.2.3 Non-Significant Trends: Besides the tendencies presented above, further tendencies which are not significant can be identified. These tendencies are shown in Fig. 14, which also uses the notation described in Fig. 11. These non-significant results point to candidate methods or practices that cannot be decided based on the available data. However, these candidate methods and practices point to further subjects worth investigating. For instance, for the practice *Model Checking*, we see a trend towards *traditional* with a *mainly agile* starting state—a trend that we have not observed in the significant results. Also, compared to the significant results, Fig. 14 makes "clear statements", i.e., we do not find one n-label, which means that the potential trends have a clear direction towards *agile* or *traditional*.

Figure 14 also fills some gaps mentioned above. For instance, while Fig. 12 lacks information about agile scaling frameworks, e.g., *LeSS*, *Nexus*, and *SAFe*, Fig. 14 provides this information. Linking this information to Fig. 10 that shows that Fisher's exact test was used for these methods due to low numbers of selections of these methods, we can conclude that these methods are either barely used or have not yet been implemented extensively in practice. Analyzing this observation requires further research.

> **Finding 9:** There are methods and practices for which we find an initial, yet not significant, indication for having an impact on the degree of agility. These candidate methods and practices warrant further research.

## 5. SUMMARY, DISCUSSION, AND ROADMAP

We summarize our key findings, which are used to steer a discussion that aims at identifying key issues in current research to develop a roadmap for future research.

### 5.1 Answering the Research Questions
The first research question defined in Section 3.1 was: *What is the degree of agility in implementing typical project disciplines in software companies?* A purely agile or traditional implementation of all project disciplines is seldom evident (Section 4.2, Finding 1b). This finding shows an even lower share than a previous study in the area of global software development that found that the number of predominantly agile projects is around 25% [50]. Finding 1a together with Fig. 8 also shows that there is a clear trend towards operating projects in an agile manner. However, the more traditional implementation of *Risk Management* indicates that agility is not implemented all the time. We argue that this points to a potential limitation of agile software development when dependable system development is the primary subject of a project. In addition, this may indicate a lack of explicit risk management techniques in agile methods.

The second research question was: *Which methods and practices influence the degree of agility of implementing the project disciplines in software companies?* Figure 10 shows that two methods and one practice are associated with a significant change of the degree of agility (using $p_{corr}$, and five methods and 15 practices with the unadjusted *p*-value), while the practices show a stronger association with changing degrees of agility (Finding 2).

Also, a subgroup of methods and practices affect the degree of agility of all project disciplines, while the *Implementation and Coding* discipline is the most affected one (Finding 3). Most practices are associated with an increase in the degree of agility (Finding 4 and 7), and once a method or practice has such an impact (high or low), this impact holds for all project disciplines (Finding 5 and 8). For instance, *Test-driven Development* is associated with a trend towards agile within all project disciplines (Fig. 12). Nevertheless, methods and practices are not exclusive, i.e., any method or practice can be found in traditional and agile software development alike. Finally, there are also methods and practices that are considered neutral with regards to the degree of agility (Finding 6, Fig. 12 and Fig. 14). Furthermore, for several methods and practices, the data does not allow for drawing final conclusions. Finding 9 and Fig. 14 summarize these methods, which indicate some trend and call for further research on their influence.

**5.2 Discussion and Roadmap**
We discuss our findings in light of current research on software processes and derive a roadmap of future research activities. We present a non-exhaustive list of topics, which emerged from discussing our findings in the context of the different research profiles of the author team. To provide some structure, the resulting topics are grouped by thematic clusters, which are provided as consecutive subsections.

*5.2.1 General Challenges in Software Engineering*
The first thematic cluster is concerned with general challenges of software and systems engineering with a particular focus on the organization of software projects. It must be noted that the topics in this section are well-known and subject to research for many years. However, especially in the context of the new application domains discussed in Section 5.2.2, these challenges are still relevant.

*Software Engineering Theory:* Aligning software engineering practices with evolving contexts is a detail-oriented and highly complex undertaking. Contexts are subject to regular change [51] and process adaptation in response to this is perhaps even more complex than some software engineers appreciate [52]. One of the conclusions from this research is that different practitioners have different views on what agility is in practice, and there is quite a significant spectrum of interpretation (findings 1, 2, 3, and 4). This leads to what might be considered an inconvenient truth for the broader community: there is insufficient theory in software engineering, this being acknowledged in earlier published material [53]. This insufficiency of theory contributes to a lack of clarity around key concepts such as agility, as highlighted in this research. Clarity is not always readily available in philosophy, and the agile manifesto [1] is certainly a philosophical artifact. But *software engineering is not a philosophy*, it is a concrete part of our day-to-day lives, we depend on it. The HELENA data research has started to create theories [42], yet, future research must put more emphasis on theorizing software engineering.

*Agile Teams in Traditional Settings:* There are specific work scenarios where agile teams have to work within more traditional structures that are influenced by traditional methods and practices [54]. It is important to understand the challenges that exist within such contexts [51], [52]. More work is necessary to understand how agile teams can work within traditional settings and in the various industry domains.

*Agile Teams in Globally Distributed Settings:* Recent research [42] suggests that hybrid organizations have in average a lower degree of agility. Most distributed software projects are neither purely agile or purely traditional in their approach to software development, but rather combine agile and traditional methods. Also, projects adopting agile scaling frameworks such as *SAFe* [55] or *LeSS* [56], nearly always employ traditional methods. Future research thus needs to provide more studies that investigate whether the mindset determines the adoption of a (purely) agile approach. This would help the community to learn whether agile is really a mindset, and shifting away from the *Waterfall*, or becoming mainly agile in any transition, is unlikely to occur unless there is a change in outlook and attitude [50]. This investigation of agile scaling frameworks and their comparison with traditional methods and "core" agile methods is necessary to determine if these scaling frameworks are really predominantly agile.

*5.2.2 Development Methods in new Contexts*
Based on the general challenges discussed above, this thematic cluster is concerned with the engineering activities as such—notably with their evolution in the context of recent technological advances.

*New Technologies:* Recent research shows that development methods have to change with emerging technologies such as machine learning (ML; [57], [58]). These changes will impact several—if not all—project disciplines. In this regard, Finding 3 is of particular interest as it shows how the different project disciplines are affected by development methods and practices, i.e., whether to develop ML applications in an agile or traditional manner. Given [59]–[61], Finding 6 becomes relevant as specific domain-relevant items can be added to a process without impacting the perception of agility. ML-research mainly focuses on the stages to build the ML model as part of software development, but methods and practices used in ML-based software development are rarely discussed. More research is required to collect, structure, and understand how software and system development methods need to change in the context of new techniques such as ML. Furthermore, in application domains such as safety-critical systems or the Internet of Things [62]– [65], companies already face pressure to become more agile. More research is necessary to understand the particularities of these domains and to design suitable methods and proper developer support for these contexts.

*Agile Model-driven Engineering:* Still, there is the open question of whether Model-driven Engineering (MDE) can be agile. These two concepts are traditionally considered incompatible, which has been questioned in previous work [66], [67]. The same discussion can be found when it comes to architecture-centric methods. Some researchers argued that architectural design does not align well with agile practices [68], [69], while others start to combine architecture- centric with agile approaches [70]. The findings 2, 4, and 7 imply that the question of a process' degree of agility is relatively independent of the method chosen. This is in line with first results on MDE-processes, indicating that agile MDE processes are feasible. These results open the potential for future research to investigate how practices can be deliberately chosen to create agile MDE-processes.

*5.2.3 Challenges in Process Design and Evolution*
While the first two thematic clusters motivate a change in the engineering of modern software-intensive systems in general, this cluster is concerned with the process as such.

*Future SPI and Agile Transitions:* In recent research, we found that Software Process Improvement (SPI) changed due to the increasing presence of agile methods. In [71], we found that (i) SPI is still conducted, but under different names, and (ii) the way SPI is conducted in the context of agile software development has changed towards the continuous learning paradigm, and also includes the various tools used in project environments [72]. In addition, many questions regarding the agile transition are left open, e.g., with respect to understanding an agile flavor of project disciplines such as requirements [73]. In this regard, findings 4, 6, and 7 are relevant in setting up and steering SPI activities. They concern the perception of project teams regarding an increased agility (Finding 4 and 7) or introduce the risk of dissatisfaction due to an unchanged degree of agility (Finding 6). Methods and practices as identified in the findings 5 and 8 provide the "low-hanging fruit". More research is required to collect, structure, and understand factors that positively or negatively affect projects and organizations, and to link these factors to methods and practices. It is imperative to understand under which conditions a high degree of agility is desirable and when a high degree of agility is counter-productive or even dangerous.

*Process Evolution:* We investigated how technology and processes co-evolve over time [74], [75]. A main finding is that technological changes are likely to change processes as well. Also, processes are changing continuously for various reasons. Especially in the context of regulated software development, awareness about evolving processes is critical, since it needs to be ensured that updated process variants remain compliant with standards [76]. Findings 5, 6, and 8, are of special importance as they enable practitioners and researchers for the first time to investigate the implications that process changes have on the degree of agility. Future research must focus on developing new prediction methods that allow practitioners to assess early on how the evolution of technologies and processes will affect agility.

*Process Deviations and Process Variants:* Process deviations and process variants seem to be common [76], [77]. We found them on the method- and practice-level. Furthermore, we know that deviations and variants can negatively impact goal achievement and teamwork [41]. Finding 2 confirms that agility is affected more strongly by practices than by methods. If a company does not reach the desired degree of agility, process deviations and variants at the level of practices should be considered as a potential cause. Future research requires more studies to identify how the degree of agility is impacted by common deviations from practice, and to develop strategies for mitigating these deviations and potential negative effects. Furthermore, new process modeling tools need to be developed that provide sophisticated design-support mechanisms, notably for the management of variants and for analyzing the effects of deviations.

*5.2.4 Human Factors in Software Engineering Practice*
Processes (Section 5.2.3) are instantiated in projects and, thus, project teams have to be enabled to undertake projects as efficiently and effectively as possible. This thematic cluster is therefore concerned with the human dimension.

*General Human Factors:* Projects and their progress are strongly affected by human and social factors [78]. Forecasts can support teams with information required to improve their performance in future iterations [79]. Especially, findings 2 and 3 highlight the influence of the used methods and practices on human perception. The chosen development methods and practices likely affect social aspects in the team such as communication behavior, which potentially influences the likelihood of social conflicts [80]. It is necessary to further study this relationship to improve the quality of forecasting instruments.

*Specialists in Software Teams:* Research on software teams composed of interdisciplinary specialists becomes more crucial as development goes agile. How this affects the process, what new tasks are emerging, and the role of team maturity are subject to research [58], [81]. Team members who are not software developers cannot be expected to find their place in the process without process change. Noll et al. [82] observed an emerging theme in literature: the original balance of scrum master, product owner and team roles are being adapted, conflated, and possibly corrupted, to suit the needs of organizations transitioning from waterfall to Scrum, or scaling Scrum to large scale organizations. Therefore, it be- comes crucial to distinguish between methods and practices that are neutral with regards to agility, and methods and practices that are not (findings 5, 6, and 8). Future research on integrating such team members will benefit from the insights in this paper by classifying observed or necessary process changes according to their impact on agility. This further leads to research on enabling teams to maintain their agility while integrating specialists into their workflow.

*Teaching Processes and Project Disciplines:* The knowledge level of students and their readiness for the software industry are subject to regular discussions among practitioners, educators, and researchers. There exist studies, e.g., [83], [84] to measure and improve the knowledge gap between software engineering education and industrial needs. Finding 3 provides us with the new insight that the different software engineering disciplines/tasks are not isolated from the question of whether development happens in an agile or traditional way. We need to investigate how teaching the software engineering disciplines, e.g., quality management or maintenance and evolution, should incorporate and address the question of how tasks change when different agile and traditional processes are applied.

# 6. CONCLUSIONS

Throughout the progress of this extensive industrial study, much consideration and reflection have been focused on the broader state of software engineering in general and this has suggested that there are some axioms that we would do well to acknowledge and work on. The fact that no one software development process or set of practices is universally and perfectly suited to all software development settings [51] should be more than just an inconvenience to those evangelists of certain processes and practices. Rather, this reality seems to reflect a basic principle: software engineering is highly applied across a very large range of situational contexts [85], [86]. Just as civil engineering uses different materials, processes and techniques for bridges, houses and roads, so too should we expect software engineers to use different tools, practices and hardware for nuclear power plants, computer games and customer relationship management. Advocating continuous software engineering [87] for nuclear reactor control software makes no more sense than evangelizing the practices for bridge builders as being perfectly suited to house builders. Therefore, we find that it is not necessarily a question of whether a context requires agility or not, it is a question of what type of agility is suitable in different contexts: which practices? Which parts of which practices? Which practices in combination?

We studied the question: *What makes agile software development agile?* To answer that question, based on 556 data points obtained in a large-scale international online survey, we studied the participants' perception of agility and whether this perception changes depending on the used methods and practices. Our findings show that (i) a clear trend towards agile software development can be observed, (ii) purely agile or traditional software development happens in rarely, (iii) most practices are associated with "being agile" and practices have a stronger impact on the perception of agility compared to methods, and (iv) there are methods and practices that are considered neutral.

However, our findings also clearly show that "agility" is in the eye of the beholder and, therefore, is a subjective concept. For instance, if people know that the base process is the *V-Model* or the *Waterfall*, everything is considered "traditional", while everything that is connected to *Scrum* or *XP* is considered "agile", regardless of whether the process of interest is the (objectively) best choice for the respective context [88]. Taking into account that the methods and practices are not used stand-alone, but to a large extent in combination with other methods and practices [8], we argue that *there is no agile software development process*. Agile software development has to be considered a *cultural topic* of project teams and software-producing organizations and how these choreograph collaboration, rather than a process modeling topic. This conclusion is supported by our finding that practices, i.e., the actual description of how the work is done, have a bigger impact than methods on the perception as to whether a process is agile.

We have presented our findings and outlined routes for future research. We consider it imperative to objectively discuss development in the context of emerging trends and technologies, which have nothing to do with *being agile or not being agile* in the first place. We argue that more emphasis must be put on process modeling and evolution to provide a meaningful methodological "backend" for the challenges in software and system development. Finally, we call for action in intensifying interdisciplinary research on (globally distributed) software teams. In a world that vitally relies on software and increasingly on remote working to produce software, as we have seen with the Covid19 pandemic, we need a clear understanding of how to organize the successful and sustainable development of high-quality software.

This research has discovered important new information about what practitioners consider agility to mean (expressed through association with various practices/methods). This has raised our sophistication of understanding of the agile concept, but, in-so-doing, has also crystallized aspects of our thinking regarding some more general realities in software engineering. The observations above could—and we suggest should—be transformed into research questions to be worked on by the broader research community.

# ACKNOWLEDGMENTS

HELENA is a huge endeavor in which many people participated and helped improve the understanding of modern software and system development. We would like to thank all persons who directly or indirectly supported HELENA. Notably, the authors would like to thank Philipp Diebold, Eckhart Hanser, Simon Oberthür, Sabrina Marczak, Guilherme Horta Travassos, Vahid Garousi, Kai Petersen, Nauman Ali, Abdur Razzak, Casper Lassenius, Nicole Novielli, Giuseppe Iaffaldano, Dan Houston, Ian De Silva, Martin Kropp, Benjamin Kanagwa, Mushtaq Raza, João Pascoal Faria, Julian Bass, Filippo Lanubile, Andreas Meier, and John Noll. Our special thanks goes to Rory V. O'Connor, who sadly left us and will be missed as a great colleague and friend.

D. Pfahl and E. Scott are partly funded by the Estonian Centre of Excellence in ICT Research (EXCITE). S. Beecham, I. Richardson, Ö. Özcan-Top, P. Clarke, and F. McCaffery were partially supported with the financial support of the Science Foundation Ireland grant 13/RC/2094 2 and co-funded under the European Regional Development Fund through the Southern & Eastern Regional Operational Programme to Lero - the Science Foundation Ireland Research Centre for Software (www.lero.ie). M. Genero, A. Vizcaino and M. Piattini research is partly funded by ECLIPSE Project (Ministerio de Ciencia, Innovación y Universidades—FEDER, RTI2018-094283-B-C31, Spain). E. Knauss has partially been funded through the Software Center research network (https://www.software-center. se) and the Vinnova FFI NGEA project (https://www.vinnova. se). R. Prikladnicki is partially funded by CNPq and Fapergs. D. Winkler and S. Biffl are supported by the Christian Doppler Research Association, the Austrian Federal Ministry for Digital and Economic Affairs and the National Foundation for Research, Technology and Development.